\documentclass[sort&compress,
    ,final            
  ]
  {aipproc}

\layoutstyle{6x9}

\begin{document}
\title{Einstein-Podolsky-Rosen-Bohm laboratory experiments:
Data analysis and simulation\footnote{FPP6 - Foundations of Probability and Physics 6, edited by A. Khrennikov et al., (AIP Conference Proceedings,
Melville and New York, in press)}}



\author{H. De Raedt}{%
 address={Department of Applied Physics, Zernike Institute for Advanced Materials,\\
          University of Groningen, Nijenborgh 4, NL-9747 AG Groningen, The Netherlands},
}

\author{K. Michielsen}{%
  address={Institute for Advanced Simulation, J\"ulich Supercomputing Centre, \\
           Research Centre J\"ulich, D-52425 J\"ulich, Germany},
}
\author{F. Jin}{%
  address={Institute for Advanced Simulation, J\"ulich Supercomputing Centre, \\
           Research Centre J\"ulich, D-52425 J\"ulich, Germany},
}

\begin{abstract}
Data produced by laboratory Einstein-Podolsky-Rosen-Bohm (EPRB) experiments is tested against the hypothesis
that the statistics of this data is given by quantum theory of this thought experiment.
Statistical evidence is presented that the experimental data,
while violating Bell inequalities, does not support this hypothesis.
It is shown that an event-based simulation model, providing a
cause-and-effect description of real EPRB experiments at a level of detail which is not covered
by quantum theory, 
reproduces the results of quantum theory of this thought experiment,
indicating that there is no fundamental obstacle for a real EPRB experiment 
to produce data that can be described by quantum theory.
\end{abstract}

\maketitle

\newcommand\sumprime{\mathop{{\sum}'}}
\newcommand\MT{Maxwell's theory}
\newcommand\QT{quantum theory}
\newcommand{\onlinecite}{\cite}
\def\DLM{DLM}
\def\DLMS{DLMs}
\def\Eq#1{(\ref{#1})}
\newcommand\ns{\,\mathrm{ns}}
\newcommand\mus{\,\mu\mathrm{s}}

\section{Introduction}\label{introduction}

In the scientific and popular literature, it is common to find statements that Einstein-Podolsky-Rosen-Bohm (EPRB)
experiments show violations of a Bell inequality and that the
experimental results~\cite{KOCH67,FREE72,CLAU78,ASPE82b,TAPS94,TITT98,WEIH98,WEIH00,WEIH07,ROWE01,FATA04,SAKA06,HNIL07}
are in favour of quantum theory~\cite{CLAU78,BELL93,BALL03}.
While there can be little doubt that the former is a firmly established fact,
it is remarkable that, at least to our knowledge, there seems to be no record
of an hypothesis test that the experimental data gathered in EPRB experiments is indeed complying
with the predictions of quantum theory for this particular experiment.

One reason for deeming such a test unnecessary may be rooted in the widespread
believe that Bell~\cite{BELL64,BELL93} would have proven that a violation of one of his inequalities
implies that the experimental findings rule out {\bf any} explanation
in terms of classical (Hamiltonian as well as non-Hamiltonian) models that satisfy Einstein's criteria for local causality.
Although the general validity of Bell's result has been questioned by many workers~\cite{%
PENA72,FINE74,FINE82,FINE82a,FINE82b,MUYN86,KUPC86,JAYN89,%
BROD93,PITO94,FINE96,KHRE99,SICA99,BAER99,%
HESS01,HESS05,ACCA05,KRAC05,SANT05,LOUB05,KUPC05,MORG06,KHRE07,ADEN07,NIEU09,MATZ09,KARL09,KHRE09,GRAF09,KHRE11,NIEU11,RAED11a},
even if the result were valid, it still remains to be shown that the experimental data
is in concert with the predictions of quantum theory.

In this paper, we report on such an hypothesis test applied to data produced
by the EPRB experiment of Weihs {\sl et al.}~\cite{WEIH98,WEIH00}.
We present compelling evidence that the experimental data,
while violating Bell inequalities, does not support the hypothesis that the data complies
with the predictions of quantum theory for the EPRB experiment.
We also demonstrate that classical, locally causal models which generate the same kind of data sets as
the laboratory EPRB experiments can reproduce the results of quantum theory,
suggesting that in the real experiments there are processes at work
which deserve to be identified and studied further.

The data sets generated by the experiment of Weihs {\sl et al.} have been
scrutinized earlier~\cite{HNIL02,HNIL07,ADEN07,BIGE09,AGUE09}.
A. Hnilo {\sl et al.} focused on ruling out certain types
of local realist models and searched for non-ergodic features
in the time-stamped data~\cite{HNIL02,HNIL07,AGUE09}.
G. Adenier and A. Khrennikov tested the hypothesis that
the data of the experiment of Weihs {\sl et al.} supports the fair sampling assumption~\cite{ADEN07}.
They showed that the relative frequencies of single-particle counts
exhibits ``non-local'' dependencies on the applied bias
and interpret this finding as evidence that the fair sampling assumption
may be violated in the experiments of Weihs {\sl et al.}~\cite{ADEN07}.
Their conclusions were criticized by J.H. Bigelow~\cite{BIGE09}
who showed that, with additional assumptions, the violation
of the non-signaling criterion, found by G. Adenier and A. Khrennikov,
can be removed by modelling the data as a
fair sample drawn from a set of sixteen coincidence counts
that obey the no-signaling criterion.

\begin{figure}[t]
\includegraphics[width=10cm]{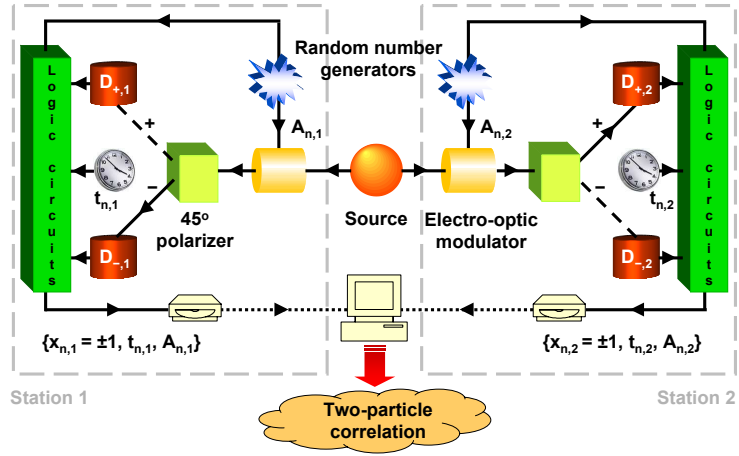}
\caption{Schematic diagram of an EPRB experiment with photons~\cite{WEIH98,WEIH00}.
}
\label{fig:Weihs}
\end{figure}

\section{EPRB laboratory experiment}\label{facts}

In Fig.~\ref{fig:Weihs}, we show a schematic diagram of the EPRB experiment with photons~\cite{WEIH98,WEIH00},
the data of which we analyze.
In this experiment, the polarization of each photon is used as the spin-1/2 degree-of-freedom in Bohm's version~\cite{BOHM51}
of the EPR gedanken experiment~\cite{EPR35}.

The source is emitting pairs of photons. The statistical properties of a large set of these photons are a priori unknown
and have to be inferred from the analysis of the recorded detection events.
As the photon arrives at station $i=1,2$ ($i=1,2$ denotes Alice's and Bob's station, respectively),
it passes through an electro-optic modulator (EOM).
The EOM in observation station $i$
rotates the polarization of the photon that passes through it
by an amount that is controlled by a voltage applied to the EOM~\cite{WEIH98,WEIH00}.
In turn, this voltage is controlled by a binary variable $A_i$, which is chosen at random~\cite{WEIH98,WEIH00}.
Optionally, a bias voltage is added to the randomly varying voltage~\cite{WEIH98,WEIH00}.
The relation between the voltage applied to the EOM and the resulting rotation of the polarization
is determined experimentally, hence there is some uncertainty in relating the applied voltage
to the rotation angle~\cite{WEIH98,WEIH00}.
Keeping in mind that the experimentally relevant variable is the applied voltage,
to simplify the presentation and the interpretation in terms of quantum theory,
we characterize the rotation of the polarization by the angle $\theta_i$, as in Refs.~\onlinecite{WEIH98,WEIH00}.
As the photon leaves the EOM, a polarizing beam splitter directs the photon to one of the two detectors,
producing a signal $x_{n,i}=\pm1$ where the subscript $n$ labels the $n$th detection event (see Fig.~\ref{fig:Weihs}).
Each station has its own clock that assigns a time-tag
to each signal generated by one of the two detectors~\cite{WEIH98,WEIH00}.
Effectively, this procedure discretizes time in intervals, the width of which is
determined by the time-tag resolution $\tau$.
In the experiment, the time-tag generators are synchronized before each run~\cite{WEIH98,WEIH00}.
The firing of a detector defines an event.
At the $n$th event at station $i$,
the dichotomic variable $A_{n,i}$ (see Fig.~\ref{fig:Weihs}),
controlling the rotation angle $\theta_{n,i}$,
the dichotomic variable $x_{n,i}$ designating which detector fires,
and the time tag $t_{n,i}$ of the detection event
are written to a file on a hard disk,
allowing the data to be analyzed long after the experiment has terminated~\cite{WEIH98,WEIH00}.
The set of data collected at station $i$ may be written as
\begin{eqnarray}
\label{Ups}
\Upsilon_i=\left\{ {x_{n,i},t_{n,i},\theta_{n,i} \vert n =1,\ldots ,N_i } \right\}
,
\end{eqnarray}
where we allow for the possibility that the number of detected events $N_i$
at stations $i=1,2$ need not (and in practice is not) to be the same
and we have used the rotation angle $\theta_{n,i}$ instead
of the corresponding dichotomic variable $A_{n,i}$ to facilitate the
comparison with the quantum theoretical description.
The data sets $\{\Upsilon_1,\Upsilon_2\}$, kindly provided to us by G. Weihs,
are the starting point for the analysis presented in this paper.

A laboratory EPRB experiment requires some criterion to decide which detection
events are to be considered as stemming from a single or two-particle system.
In EPRB experiments with photons, this decision is taken on the basis of coincidence in time~\cite{WEIH98,CLAU74}.
Here we adopt the procedure employed by Weihs {\sl et al.}~\cite{WEIH98,WEIH00}.
Coincidences are identified by comparing the time differences
$t_{n,1}-t_{m,2}$ with a window $W$~\cite{WEIH98,WEIH00,CLAU74},
where $n=1,\ldots,N_1$ and $m=1,\ldots,N_2$.
By definition, for each pair of rotation angles $a$ and $b$,
the number of coincidences between detectors $D_{x,1}$ ($x =\pm $1) at station 1 and
detectors $D_{y,2}$ ($y =\pm $1) at station 2 is given by
\begin{eqnarray}
\label{Cxy}
C_{xy}=C_{xy}(a,b)&=&
\sum_{n=1}^{N_1}
\sum_{m=1}^{N_2}
\delta_{x,x_{n ,1}} \delta_{y,x_{m ,2}}
\delta_{a ,\theta_{n,1}}\delta_{b,\theta_{m,2}}
\Theta(W-\vert t_{n,1} -t_{m ,2}\vert)
,
\end{eqnarray}
where $\Theta (t)$ is the Heaviside step function.
In Eq.~(\ref{Cxy}) the sum over all events has to be carried out such that each event (= one detected photon) contributes only once.
Clearly, this constraint introduces some ambiguity in the counting procedure as there is a priori, no clear-cut criterion
to decide which events at stations $i=1$ and $i=2$ should be paired.
One obvious criterion might be to choose the pairs such that $C_{xy}$ is maximum but as we explain later,
such a criterion renders the data analysis procedure (not the data production!) acausal.
It is trivial though (see later) to analyse the data generated by the experiment of Weihs {\sl et al.}
such that conclusions do not suffer from this artifact.
In general, the values for the coincidences
$C_{xy}(a,b)$ depend on the time-tag resolution $\tau$
and the window $W$ used to identify the coincidences.

The single-particle averages and correlation between the coincidence counts
are defined by
\begin{eqnarray}
\label{Exy}
E_1(a,b)&=&
\frac{\sum_{x,y=\pm1} x C_{xy}}{\sum_{x,y=\pm1} C_{xy}}
=
\frac{C_{++}-C_{--}+C_{+-}-C_{-+}}{C_{++}+C_{--}+C_{+-}+C_{-+}}
\nonumber \\
E_2(a,b)&=&
\frac{\sum_{x,y=\pm1} yC_{xy}}{\sum_{x,y=\pm1} C_{xy}}
=
\frac{C_{++}-C_{--}-C_{+-}+C_{-+}}{C_{++}+C_{--}+C_{+-}+C_{-+}}
\nonumber \\
E(a,b)&=&
\frac{\sum_{x,y=\pm1} xy C_{xy}}{\sum_{x,y=\pm1} C_{xy}}
=
\frac{C_{++}+C_{--}-C_{+-}-C_{-+}}{C_{++}+C_{--}+C_{+-}+C_{-+}}
,
\end{eqnarray}
where the denominator $N_c=C_{++}+C_{--}+C_{+-}+C_{-+}$
in Eq.~(\ref{Exy}) is the sum of all coincidences.
In practice, coincidences are determined by a four-step procedure~\cite{WEIH00}:
\begin{enumerate}
\item{Compute a histogram of time-tag differences $t_{n,1}-t_{m,2}$ of pairs of detection events.}
\item{Determine the time difference $\Delta_G$ for which this histogram shows a maximum.}
\item{Add $\Delta_G$ to the time-tag data of Alice, thereby moving the position of the maximum of the histogram to zero.}
\item{Determine the coincidences using the new time-tag differences,
each photon contributing to the coincidence count at most once.}
\end{enumerate}
The global offset, denoted by $\Delta_G$, may be attributed
to the loss of synchronization of the clocks used in the stations of Alice and Bob~\cite{WEIH00}.

\subsection{Role of the time window and acausal data processing}\label{role}

Most theoretical, local-realistic treatments of the EPRB experiment assume that the correlation,
as measured in the experiment, is given by~\cite{BELL93}
\begin{eqnarray}
\label{CxyBell}
C_{xy}^{(\infty)}(a,b)&=&\sum_{n=1}^N\delta_{x,x_{n ,1}} \delta_{y,x_{n ,2}}
\delta_{a ,\theta_{n,1}}\delta_{b,\theta_{m,2}}
,
\end{eqnarray}
which is obtained from Eq.~(\ref{Cxy}) by taking the limit $W\rightarrow\infty$
and omitting events such that $N=N_1=N_2$.
A naive argument that might justify taking this limit is
the hypothesis that for ideal experiments, the value of $W$ should not matter
because the time window only serves to identify pairs.
However, this argument does not apply to real experiments:
The analysis of the data of the experiment of Weihs {\sl et al.} shows that
the average time between pairs of photons is of the order of $30\mus$ or more,
much larger than the typical values (of the order of a few nanoseconds)
of the time-window $W$ used in the experiments~\cite{WEIH00}.
In other words, in practice, the identification of photon pairs
does not require the use of $W$'s of the order of a few nanoseconds in which case
the use of a time-coincidence window does not create a ``loophole''.
A narrow time window mainly acts as a filter that selects pairs,
the photons of which experienced time delays that differ by the order of nanoseconds.

The use of a global offset $\Delta_G$, determined
by maximizing the number of coincidences, introduces an element of non-causality in the
analysis of the experimental results (but not in the original data itself):
Whether or not at a certain time, a pair contributes to the number of coincidences depends
on {\bf all} the coincidences, also on those at later times.
This is an example of an acausal filtering~\cite{PRES03}:
The output (coincidence or not) depends on both all previous
and all later inputs (the detection events and corresponding time-tags).
Therefore, data analysis of EPRB experiments that employs a
global offset $\Delta_G$ to maximize the number of coincidences,
is intrinsically acausal:
The notion of coincidences happening inside or outside the light cone becomes irrelevant.

In spite of this caveat, very useful information can be extracted from
data sets produced by the experiment of Weihs {\sl et al.}, the reason being
that the results of the analysis become independent of $\Delta_G$ if the time-window
$W$ is taken to be sufficiently large (but much smaller than the average time between successive events).
As it is our aim to test whether the data of the experiment of Weihs {\sl et al.} comply with quantum theory,
not to find the maximum violation of some inequality,
we will not dwell on this issue any further and simply discard conclusions that depend
on the use of a non-zero $\Delta_G$.

\section{hypothesis test}\label{test}

According to quantum theory of the EPRB thought experiment, the results of repeated measurements
of the system of two $S=1/2$ particles in the spin state $|\Phi\rangle$
are given by the single-spin expectation values
\begin{eqnarray}
\widehat E_1({a})&=&\langle \Phi|\mathbf{S}_1\cdot\mathbf{a}|\Phi\rangle=\langle \Phi|\mathbf{S}_1|\Phi\rangle\cdot\mathbf{a}
,
\nonumber \\
\widehat E_2({b})&=&\langle \Phi|\mathbf{S}_2\cdot\mathbf{b}|\Phi\rangle=\langle \Phi|\mathbf{S}_2|\Phi\rangle\cdot\mathbf{b}
,
\label{Ei}
\end{eqnarray}
and the two-particle correlations
%
$
\widehat E({a},{b})=
\langle \Phi|\mathbf{S}_1\cdot\mathbf{a}\; \mathbf{S}_2\cdot\mathbf{b}|\Phi \rangle
=\mathbf{a}\cdot\langle \Phi|\mathbf{S}_1\; \mathbf{S}_2|\Phi \rangle \cdot\mathbf{b}
$
where $\mathbf{a}=(\cos a,\sin a)$ and $\mathbf{b}=(\cos b,\sin b)$ specify
the directions of the analyzers (corresponding to the rotations of the polarization due to the EOM's).
We have introduced the notation $\widehat{\phantom{E}}$
to distinguish the quantum theoretical prediction from the results
obtained by analysis of the experimental data sets.

Quantum theory of the EPRB thought experiment assumes that $|\Phi\rangle$ does
not depend on ${a}$ or ${b}$.
Therefore, from Eq.~(\ref{Ei}) it follows immediately that
$\widehat E_1({a})$ does not depend on ${b}$
and that
$\widehat E_2({b})$ does not depend on ${a}$.
Under the hypothesis that quantum theory describes the
data collected in the laboratory EPRB experiment,
we may expect that
$E_1({a},{b})\approx\widehat E_1({a})$
and
$E_2({a},{b})\approx\widehat E_2({b})$,
exhibit the same independencies.
This is the basis of our test.

In practice, we are dealing with real data and not with mathematical expressions
such as Eq.~(\ref{Ei})
and therefore, we need a criterion to decide whether or not the data
is in agreement with the value predicted by quantum theory of the EPRB experiment.
In line with standard statistical hypothesis testing, we adopt the following criterion:
\begin{center}
\framebox{
\parbox[t]{0.92\hsize}{%
The data for $E_1(a,b)$ ($E_2(a,b)$) are considered to be in conflict with
the prediction of quantum theory of the EPRB experiment if it shows a dependency
on $b$ ($a$) that exceeds five times the upper bound $1/\sqrt{N_c}$ to the standard deviation $\sigma_{N_c}$.
}
}
\end{center}

\noindent
Data that does not satisfy our criterion exhibits a spurious kind of ``non-locality''
that {\bf cannot} be explained by the quantum theoretical description of the EPRB experiment.

A key feature of our test is that it does not rely on any particular property of the state $|\Phi\rangle$.
For instance, if in a laboratory EPRB experiment we find that
$E_1(a,b)$ shows a dependence on $b$
that exceeds five times the standard deviation,
this dependence {\bf cannot} be attributed to
$|\Phi\rangle$ deviating from the singlet state.

%
\begin{figure}[t]
\centering
\includegraphics[width=7.5cm]{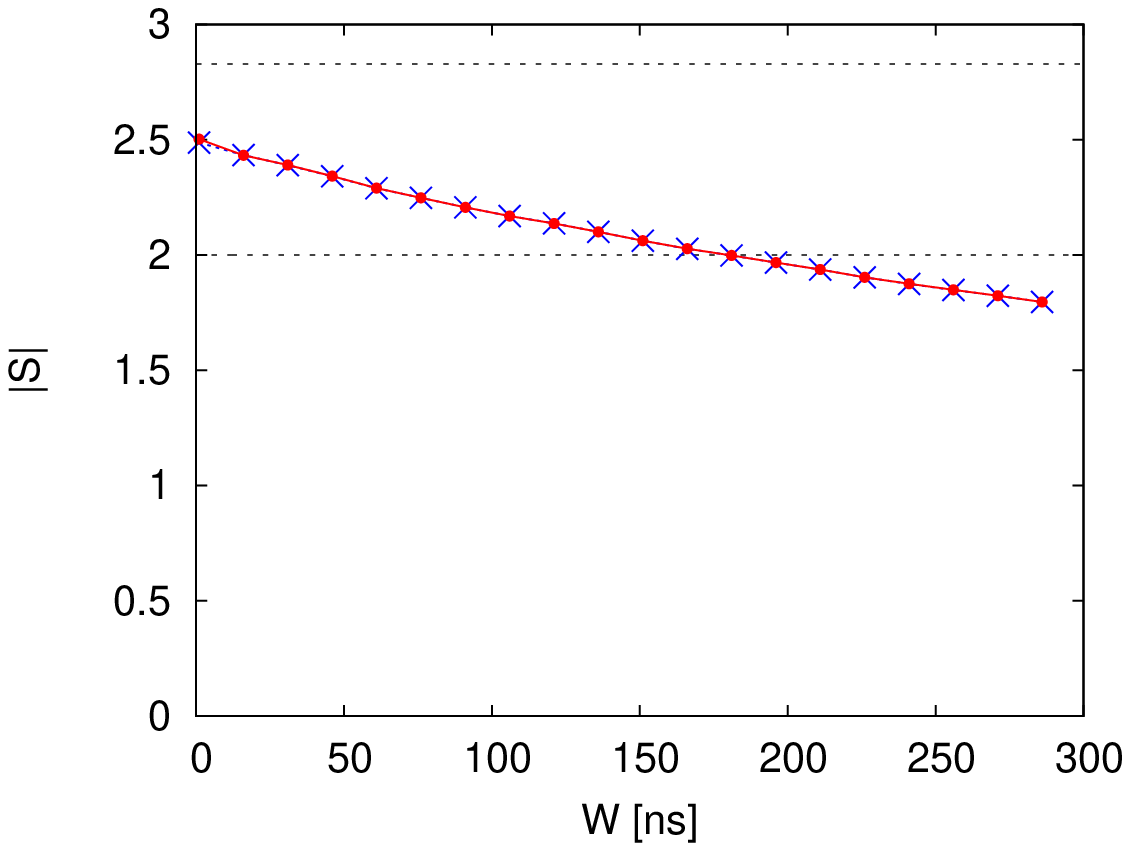}
\includegraphics[width=7.5cm]{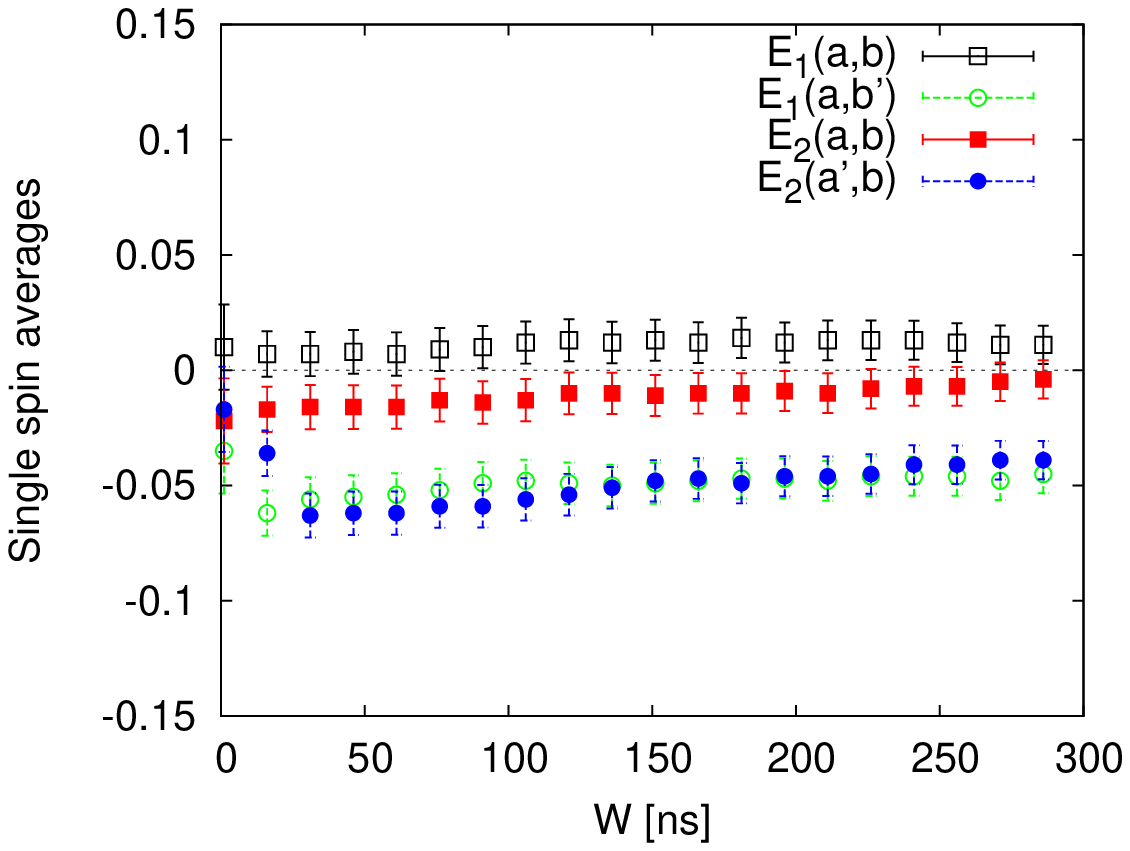}
\caption{
Analysis of the data set {\bf newlongtime2}.
Left:
$|S|=|E(a,b)-E(a,b')+E(a',b)+E(a',b')|$ as a function of the time window $W$ and
$a=0$, $a'=\pi/4$, $b=\pi/8$ and $b'=3\pi/8$.
The dashed lines represent the maximum value for a quantum system of two $S=1/2$ particles
in a separable (product) state ($|S|=2$) and in a singlet state ($|S|=2\sqrt{2}$), respectively.
Blue crosses: $\Delta_G=0$.
Red bullets connected by the red solid line: $\Delta_G=0.5\ns$.
Right:
Selected single-particle averages
as a function of $W$ and for $\Delta_G=0$.
The error bars correspond to $2.5$ standard deviations.
For small $W$, the total number of coincidences
is too small to yield statistically meaningful results.
For $W>20\ns$ the change of some of these single-spin averages
observed by Bob (Alice) when Alice (Bob) changes her (his) setting,
systematically exceeds five standard deviations,
suggesting it is highly unlikely that the data is in
concert with quantum theory of the EPRB experiment.
}
\label{fig.A5}
\end{figure}

Figure~\ref{fig.A5}(left) shows the typical results of the Bell function $S$ as a function of $W$.
For $W<150\ns$, the Bell inequality $|S|\le2$ is clearly violated.
For $W>200\ns$, much less than the average time ($>30\mus$) between two coincidences,
the Bell inequality $|S|\le2$ is satisfied, demonstrating that the ``nature''
of the emitted pairs is not an intrinsic property of the pairs themselves but also depends on
the choice of $W$ made by the experimenter.
For $W>20\ns$, there is no significant statistical evidence that the ``noise'' on the data depends on $W$
but if the only goal is to maximize $|S|$, it is expedient to consider $W<20\ns$.
From these experimental results, it is clear that the time-coincidence window does not constitute a ``loophole''.
Not only is it an essential element of these EPRB experiments,
it also acts as a filter that is essential for the data to violate a Bell inequality.

Figure~\ref{fig.A5}(right) shows results of a selection of single-particle expectations as a function of $W$.
It is clear that these results violate our criterion for being compatible
with the prediction of quantum theory of the EPRB model.
According to standard practice of hypothesis testing, the likelyhood that this data set
can be described by the quantum theory of the EPRB experiment should be considered as extremely small.
This finding is not an accident:
Our analysis of 23 data sets produced by the experiment of Weihs {\sl et al.}
shows that none of these data sets satisfies our hypothesis test for being compatible with
the predictions of quantum theory of the EPRB model.

\begin{table}[t]
\begin{tabular}{cccccccccc}
\hline
    {$r_1$}
  & {$r_2$}
  & {$\widehat E_1({a})$}
  & {$\widehat E_1({a'})$}
  & {$\widehat E_2({b})$}
  & {$\widehat E_2({b'})$}
  & {$\widehat E({a},{b})$}
  & {$\widehat E({a},{b'})$}
  & {$\widehat E({a'},{b})$}
  & {$\widehat E({a'},{b'})$}
  \\
\hline
$+0.17$ & $-0.01$ & $-0.17$ & $-0.23$ & $+0.11$ & $-0.13$ & $-0.72$ & $+0.47$ & $-0.52$ & -- \\
$-0.06$ & $-0.01$ & $+0.06$ & $+0.03$ & $-0.05$ & $-0.03$ & $-0.69$ & $+0.45$ & --      & $-0.71$ \\
$+0.17$ & $-0.16$ & $-0.29$ & $-0.32$ & $+0.27$ & $+0.31$ & $-0.83$ & --      & $-0.62$ & $-0.87$ \\
$-0.06$ & $-0.17$ & $+0.13$ & $-0.08$ & $+0.16$ & $+0.13$ & --      & $+0.46$ & $-0.50$ & $-0.72$ \\
\hline
\end{tabular}
\caption{Results of the numerical solution of the set of three groups of equations such as Eq.~(\ref{de0})
for data taken from the experimental data set
{\bf newlongtime2} at $W=50\ns$.
The pair of settings that has not been included corresponds to the entry indicated with the --.
For instance, the results of the first row have been obtained by excluding the data
for the setting $(a',b')$.
Consistency demands that the numbers in each column are close to each other.
The fact that the four solutions are inconsistent suggests that the data is incompatible
with quantum theory for the EPRB thought experiment.
}
\label{tab:a}
\end{table}

\section{Role of the detection efficiency}\label{detection}

For the experimental set up of Weihs {\sl et al.},
the dependence of $E_1(a,b)$ ($E_2(a,b)$) on ${b}$ (${a}$) cannot
be attributed to detection efficiencies of the detectors at station (1),
assuming the latter as constant during the time the data is collected.
This can be seen as follows.

As photons may be absorbed when passing through the EOM and
as detectors do not register all incident photons, we may write
\begin{eqnarray}
C_{xy}(a,b)&=&\kappa_1(a)\kappa_2(b)\eta_1(x)\eta_2(y)N P(xy|ab)
,
\end{eqnarray}
where $0<\kappa_i(.)\le1$ and $0<\eta_i(.)\le1$
represent the efficiency of the EOM and detectors at station $i=1,2$ respectively,
$N$ is the number of photon pairs emitted by the source
and $P(xy|ab)=(1+x\widehat E_1({a})+y\widehat E_2({b})+xy\widehat E({a},{b}))/4$
is the most general form of the probability for a pair $x,y=\pm1$, compatible with quantum theory of the EPRB thought experiment.
Note that the experiment of Weihs {\sl et al.} uses polarizing beam splitters, and therefore the
detectors receive photons with fixed polarization.
Hence the detection efficiencies $\eta_1(x)$ and $\eta_2(y)$ are not expected
to depend on the polarization of the photons that leave the EOMs~\cite{WEIH00}.
After some elementary algebra, we find
\begin{eqnarray}
E_1(a,b)&=&
\frac{r_1+\widehat E_1({a})+r_1r_2\widehat E_2({b})+r_2\widehat E({a},{b})}{
1+r_2\widehat E_1({a})+r_1\widehat E_2({b})+r_1r_2\widehat E({a},{b})}
.
\nonumber \\
E_2(a,b)&=&
\frac{r_2+r_1r_2\widehat E_1({a})+\widehat E_2({b})+r_1\widehat E({a},{b})}{
1+r_2\widehat E_1({a})+r_1\widehat E_2({b})+r_1r_2\widehat E({a},{b})}
,
\nonumber \\
E(a,b)&=&
\frac{r_1r_2+r_2\widehat E_1({a})+r_1\widehat E_2({b})+\widehat E({a},{b})}{
1+r_2\widehat E_1({a})+r_1\widehat E_2({b})+r_1r_2\widehat E({a},{b})}
,
\label{de0}
\end{eqnarray}
where $-1\le r_i=(\eta_i(+1)-\eta_i(-1))/(\eta_i(+1)+\eta_i(-1))\le 1$
parameterizes the relative efficiencies of the detectors at stations $i=1,2$.
The parameters $r_1$, $r_2$ and the state of the quantum system (fully described by
$\widehat E_1({a})$, $\widehat E_2({b})$, $\widehat E({a},{b})$)
can be determined by considering the set of nine equations for the
three pairs of settings.
Taking for instance the data for the set of settings $\{(a,b),(a,b'),(a',b)\}$,
yields three times three equations of the form Eq.~(\ref{de0})
with nine unknowns which are easily solved numerically.
As an illustrative example, we examine the data at $W=50\ns$ taken from the experimental data set
{\bf newlongtime2}, see also Fig.~\ref{fig.A5}.
The results of the numerical solution of the nine equations
is given by the first row of Table~\ref{tab:a}.
Similarly, we can construct three additional sets of nine equations.
Their solutions are also presented in Table~\ref{tab:a}.
Clearly, there is no way in which these four solutions can be considered as compatible.
Needless to say, the full set of 12 equations has no solution at all.
This incompatibility is not accidental, it is generic.
Using a different approach of analyzing the data, J.H. Bigelow
came to a similar conclusion~\cite{BIGE09}.
Apparently, including a model for the detector efficiency does not resolve the conflict
between the experimental data of Weihs {\sl et al.} and quantum theory of the EPRB thought experiment.

\section{Discrete event simulation}\label{model}\label{simulationmodel}

The fact that the data produced by the EPRB experiment of Weihs {\sl et al.} cannot be brought in harmony with
the predictions of quantum theory for this experiment raises the question whether there exist
(simulation) models of this experiment that produce the same kind of data sets as the real experiment and
do not rely on concepts of quantum theory, yet reproduce the results of quantum theory of the EPRB experiment.

The possibility that such models exist was, to our knowledge, first pointed out by A. Fine~\cite{FINE82}.
The key of Fine's so-called synchronization model is the use of a filtering mechanism, essentially
the time-coincidence window employed in laboratory EPRB experiments.
Fine pointed out that such a filtering mechanism may lead to violations
of the inequality $|S|\le2$, opening the route to a description in terms of locally causal, classical models.
A concrete model of this kind was proposed by S. Pascazio who showed
that his model approximately reproduces the correlation of the singlet state~\cite{PASC86}
with an accuracy that seems far beyond what is experimentally accessible to date.
Later, Larson and Gill showed that Bell-like inequalities need to be modified
in the case that the coincidences are determined by a time-window filter~\cite{LARS04}.
Finally, a time-tag model that exactly reproduces the results of quantum theory for the singlet
and uncorrelated state was found~\cite{RAED06c,RAED07b,ZHAO08,MICH11a}.

A minimal, discrete-event simulation model of the EPRB experiment by Weihs {\sl et al.} (see Fig.~\ref{fig:Weihs})
requires a specification of the information carried by the particles,
of the algorithm that simulates the source and
the observation stations, and of the procedure to analyze the data.

\begin{itemize}
\item{{\bf Source and particles}:
The source emits particles that carry a vector
${\bf S}_{n,i}=(\cos(\xi_{n}+(i-1)\pi/2) ,\sin(\xi_{n}+(i-1)\pi/2))$,
representing the polarization of the photons.
This polarization is completely characterized by $\xi _{n}$ and
the direction $i=1,2$ to which the particle moves.
A uniform pseudo-random number generator is used to pick the angle $0\le\xi _{n}<2\pi$.
Clearly, the source emits two particles with a mutually orthogonal, hence correlated but otherwise
random polarization.
}
\item{{\bf EOM}:
The EOM rotates the polarization of the incoming particle by an angle $\zeta$,
that is $\alpha_{n,i}\equiv\mathrm{EOM}_i(\xi_{n}+(i-1)\pi/2,\zeta_i)=\xi_{n}+(i-1)\pi/2-\zeta_i$ symbolically.
Mimicking the experiment of Weihs {\sl et al.},
we generate two binary uniform pseudo-random numbers $A_i=0,1$ and use them
to choose the value of the angles $\zeta_i$, that is
$\zeta_1=a(1-A_1)+A_1(a+\pi/4)$ and $\zeta_2=b(1-A_2)+A_2(b+\pi/4)$.
}
\item{{\bf Polarizing beam splitter}:
The simulation model for a polarizing beam splitter is defined by the rule
\begin{eqnarray}
x_{n,i}=\left\{
\begin{array}{lll}
1 & \mbox{if} & r_n\le \cos^2\alpha_{n,i}\\
0 & \mbox{if} & r_n > \cos^2\alpha_{n,i}
\end{array}
\right.
,
\label{sg1}
\end{eqnarray}
where $0< r_n<1$ are uniform pseudo-random numbers.
It is easy to see that for fixed polarization $\alpha_{n,i}=\alpha_i$, this rule generates
events such that
\begin{eqnarray}
\lim_{N\rightarrow\infty}
\frac{1}{N}\sum_{n=1}^N x_{n,i} = \cos^2\alpha_{n,i}
,
\label{sg2}
\end{eqnarray}
with probability one,
showing that the distribution of events complies with Malus law.
}
\item{{\bf Time-tag model}: The time-tag model of Pascazio is based on the assumption
that the interaction between the photon and the detector follows an exponential
probability law~\cite{PASC86}.
This model assumes that as the photon enters the detector at time $t=0$,
it generates an electrical signal between $t$ and $t+dt$ with probability
$p(t)dt=\lambda(\alpha_{n,i})e^{-\lambda(\alpha_{n,i})t}dt$~\cite{PASC86}
or, in the notation used in the present paper,
that the time tag is given by
\begin{eqnarray}
t_{n,i}&=&-\lambda(\alpha_{n,i}) \ln r
,
\label{sg3a}
\end{eqnarray}
where $\lambda(\alpha_{n,i})$ is the inverse ``life-time'' of the photon-detector interaction
which is assumed to depend on the polarization $\alpha_{n,i}$~\cite{PASC86},
and $r$ is a uniformly distributed random number (which changes with each event).

As is well-known, as light passes through an EOM (which is essentially a tuneable wave plate), it experiences a retardation
depending on its initial polarization and the rotation by the EOM.
On the level of individual particles, in our time-tag model, we hypothesize that this delay is
represented by the time tag
\begin{eqnarray}
t_{n,i}&=&\lambda(\alpha_{n,i}) r
,
\label{sg3b}
\end{eqnarray}
that is, the time tag is distributed uniformly over the interval $[0, \lambda(\alpha_{n,i})]$~\cite{RAED06c}.
It has been shown that for $\lambda(\alpha_{n,i})=T_0\sin^4 2\alpha_{n,i}$
this model, in combination with the model
of the polarizing beam splitter, exactly reproduces the results
of quantum theory of the EPRB experiments in the limit $W\rightarrow0$~\cite{RAED07b,ZHAO08}.
We therefore adopt the expression $\lambda(\alpha_{n,i})=T_0 \sin^4 2\alpha_{n,i}$ for
both time-tag models, leaving only $T_0$ as an adjustable parameter.
}
\item{{\bf Data analysis}:
The simulation algorithm generates the data sets $\Upsilon_i$,
just as experiment does.
We analyse these data sets in exactly the same manner as in the case
of the experimental data.
}
\end{itemize}

\begin{figure}[t]
\centering
\includegraphics[width=7.5cm]{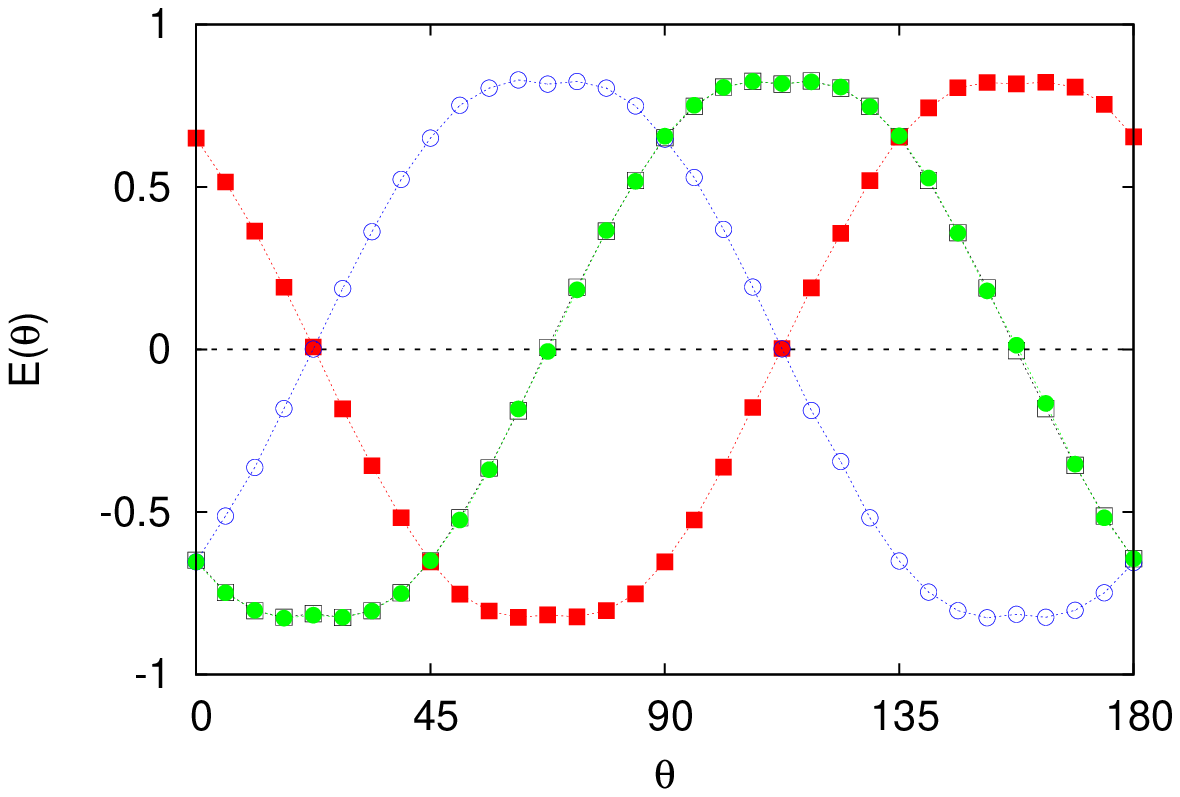}
\includegraphics[width=7.5cm]{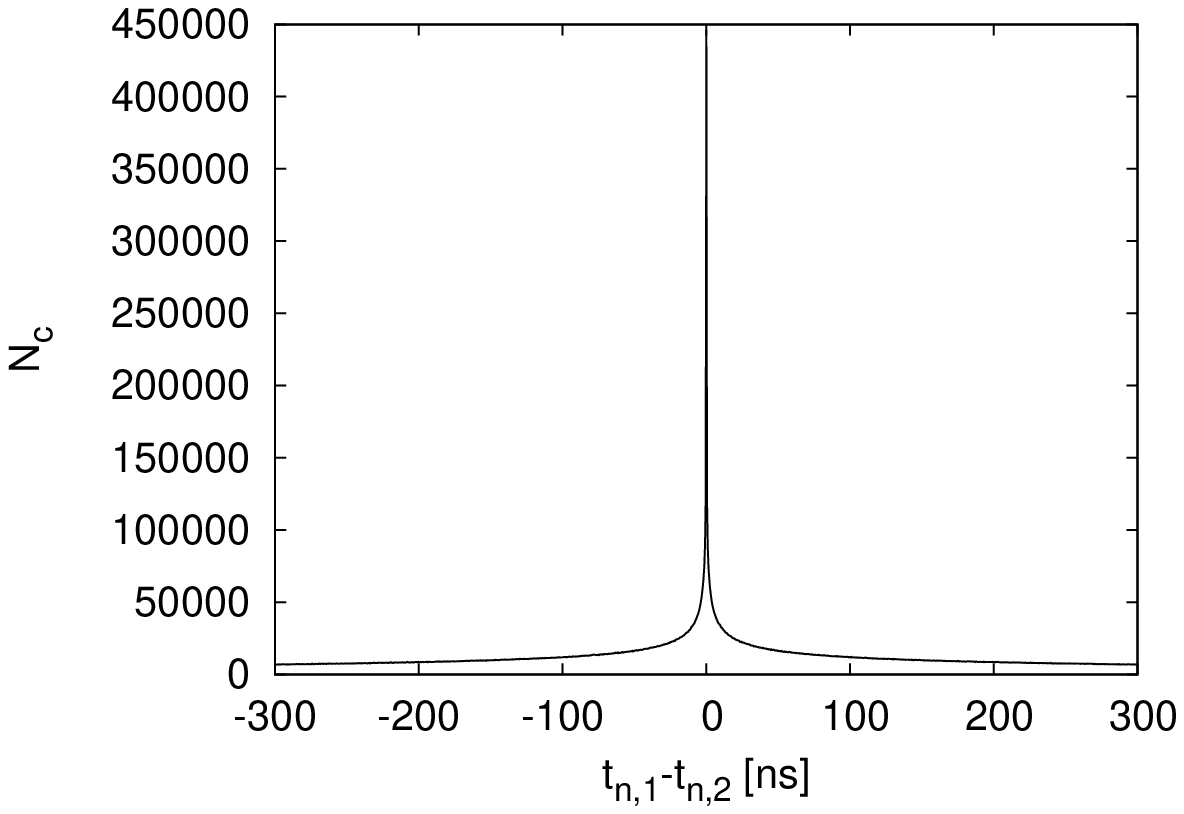}
\caption{
Simulation results produced by the locally causal, discrete-event simulation
model using the time-tag model Eq.~(\ref{sg3b}).
Left: The two-particle correlation as a function of the angle $\theta$
for $W=50\ns$.
Open squares:
$E(\theta)=E(\theta,\pi/8)$;
Open Circles:
$E(\theta)=E(\theta+\pi/4,\pi/8)$;
Solid squares:
$E(\theta)=E(\theta,3\pi/8)$;
Bullets:
$E(\theta)=E(\theta+\pi/4,3\pi/8)$;
The dashed lines are guides to the eye only.
Right:
Histograms of the differences of the time-tags.
The total number of pairs generated is $10^6$ and $T_0=2000\ns$.
}
\label{fig.S0}
\end{figure}

This algorithm fully complies with
Einstein's criterion of local causality on the ontological level: Once the particles
leave the source, an action at observation station 1 (2) can, in no way,
have a causal effect on the outcome of the measurement at observation
station 2 (1).


In Fig.~\ref{fig.S0} we present some typical simulation results
for the two-particle correlation $E(a,b)$ and the distribution
of time-tag differences, as obtained by using time-tag model Eq.~(\ref{sg3b}).
The single-particle averages $E_1(a)$ and $E_2(b)$ (data not shown) are zero up to
the usual statistical fluctuations and do not show any statistically
relevant dependence on $b$ or $a$, respectively,
in concert with a rigorous probabilistic treatment of this simulation model~\cite{ZHAO08}.

For $W=2\ns$, the results for $E(a,b)$ fit very well to the prediction of quantum theory for the EPRB experiment.
From these data (not shown), we extract $|S|=2.89$ and $|S|=2.82$ for Pascazio's and our time-tag model respectively.
Note that for both models, the maximum value of $|S|$ is not limited by the quantum theoretical result
and can, depending on the choice of $\lambda(\alpha_{n,i})$, reach values that are very close to 4~\cite{RAED07c}.
For $W=50\ns$, the two models yield $|S|=2.63$ and $|S|=2.62$, respectively.
Both results compare very well with the values extracted from the experimental data of Weihs {\sl et al.},
which for $W=50\ns$, are between 2 and 2.57.
In all cases, the distribution of time-tag differences is sharply peaked and displays long tails,
in qualitative agreement with experiment~\cite{WEIH00}.

\section{Conclusion}\label{conclusion}

It is highly unlikely that quantum theory describes the
data of the EPRB experiment that we have analyzed.
This suggests that in the real experiment, there may be processes at work which have not been identified yet.
On the other hand, event-based simulation models provide a
cause-and-effect description of real EPRB experiments at a level of detail which is not covered
by quantum theory, such as the effect of the choice of the time-window.
Some of these simulation models exactly reproduce the results of quantum theory of the EPRB experiment,
indicating that there is no fundamental obstacle for an EPRB experiment
to produce data that can be described by quantum theory.
In any case, the popular statement that EPRB experiments agree with quantum theory
does not seem to have a solid scientific basis yet.

\section{Acknowledgment}
We thank G. Weihs for providing us with the data sets of their EPRB experiments
and S. Reuschel for help during the initial phase of the present work.
We have profited from discussions with G. Adenier, J.H. Bigelow, K. De Raedt, K. Hess, A. Khrennikov, S. Reuschel, and G. Weihs.
This work is partially supported by NCF, the Netherlands.

\bibliographystyle{aipproc}       
\bibliography{/d/papers/epr11}   

\end{document}